\documentclass[aps,prl,twocolumn,superscriptaddress, showpacs]{revtex4}
\usepackage{graphicx}
\begin{document}

\title{Microtubule dynamics depart from wormlike chain model}
\author{Katja M. Taute}
\affiliation{Center for Nonlinear Dynamics, University of Texas at
  Austin, 1 University Station C1610, Austin TX 78712, U.S.A.}

\author{Francesco Pampaloni}
\affiliation{Cell Biology and Biophysics Unit, European Molecular
  Biology Laboratory, Meyerhofstra\ss{}e 1, 69117 Heidelberg, Germany}

\author{Erwin Frey}
\affiliation{Arnold Sommerfeld Center for Theoretical Physics and
  CeNS, Department of Physics, Ludwig-Maximilians-Universit\"at
  M\"unchen, Theresienstra\ss{}e 37, D-80333 M\"unchen, Germany}

\author{Ernst-Ludwig Florin}
\email{florin@chaos.utexas.edu}
\affiliation{Center for Nonlinear Dynamics, University of Texas at
  Austin, 1 University Station C1610, Austin TX 78712, U.S.A.}

\date{\today}

\begin{abstract}
Thermal shape fluctuations of grafted microtubules were studied using high
resolution particle tracking of attached fluorescent beads. First mode
relaxation times were extracted from the mean square displacement in
the transverse coordinate. 
For microtubules shorter than $\sim$$10\,\mu$m, the relaxation
times were found to follow an $L^2$ dependence instead of $L^4$ as
expected from the standard wormlike chain model. 
This length dependence is shown to result from a complex length dependence of
the bending stiffness which can be understood as a result of the molecular architecture of
microtubules. 
For microtubules shorter than $\sim$$5\,\mu$m, high drag coefficients
indicate contributions from internal friction to the fluctuation dynamics. 
\end{abstract}

\maketitle
Microtubules are cytoskeletal protein filaments that play an
essential role in a multitude of cell functions in all eucaryotes. 
While specialized structures such as cilia, flagella
and axons require microtubule lengths of up to several hundred
micrometers, fundamental tasks like cell division and intracellular
transport involve microtubules with lengths that are comparable to or
 smaller than typical eucaryotic cell sizes (10-20$\,\mu$m). 
In solution, microtubules undergo strong thermal
shape fluctuations but, due to their rigidity, maintain an average
direction and are therefore referred to as semiflexible. 
The standard model for semiflexible polymers is the wormlike chain
\cite{KratkyPorod, Saito} which envisions a homogeneous,
isotropic, continuously flexible rod characterized by its bending
stiffness,~$\kappa$.

Recently, evidence has accumulated suggesting that the
wormlike chain as the standard model for semiflexible polymers may
constitute an oversimplification in the case of microtubules due to
their highly anisotropic molecular architecture \cite{kis, fra}.
Microtubules consist of strings of $\alpha$- and
$\beta$-tubulin heterodimers, so-called protofilaments, that are
arranged in parallel forming a hollow tube of 25$\,$nm external diameter.
Neighboring protofilaments are shifted relative to each other, giving
rise to helicity. 
Both the rise per turn and the number of protofilaments
may vary, however 3 monomers per turn and 13 constituent
filaments are most common \textit{in vivo} \cite{Sept, defects}. 
This protofilament architecture makes microtubules a model system for
the generalized theory of wormlike bundles \cite{clausPRL, bathe}, which in
addition to various other biological examples also describes the
mechanics of carbon nanotube
bundles.

Few studies have addressed the dynamics of
thermal shape fluctuations of microtubules even though the
timescales of these fluctuations affect the timescales on which biological
functions occur.
Caspi et al.~\cite{Caspi} examined the transverse mean square displacement of microtubules in
networks, finding a power law behavior $t^{3/4}$.
Janson and Dogterom~\cite{Janson} extracted autocorrelation times for several modes from a shape
analysis of growing grafted microtubules, while 
Brangwynne et al.~\cite{bend} applied this procedure to stabilized
fluorescent microtubules. 
For technical reasons, all of these studies restricted themselves to lengths of
several tens of micrometers, while microtubule lengths relevant to
cell functions such as cell division are much shorter.
In a study of equilibrium position distributions \cite{fra}, the most significant
systematic deviations from the wormlike chain model were found for
microtubules shorter than $\sim$$20\,\mu$m which coincides with the
length regime most crucial to cell division and intracellular transport.
This suggests that, in this length regime, corresponding deviations should exist for dynamical parameters
such as the relaxation time.

In this letter we present an analysis of first mode relaxation times
for microtubules of length 2-30$\,\mu$m. 
Microtubules are grafted to a substrate at one end and data is
extracted from the mean square displacement of the
transverse coordinate. 
Capturing dynamics for microtubules that are only several $\mu$m long
is challenging since the corresponding spatial and temporal
scales become increasingly small. 
We overcome these difficulties by using small fluorescent beads
attached to the microtubules as tracer particles. Their position can
be tracked at frame rates of up to 30$\,$Hz using a high quantum efficiency CCD camera.
In contrast to shape analysis techniques that provide low resolution
(tens of nanometers \cite{Janson, bend}) position data for the whole filament,
the use of the tracer bead yields spatial information with a
precision of a few nanometers for one specific point on the filament's
contour.
Standard semiflexible polymer models are then used to infer relaxation
times, bending stiffnesses and drag coefficients.

\begin{figure}
  \centering
  \includegraphics[width=8.5cm]{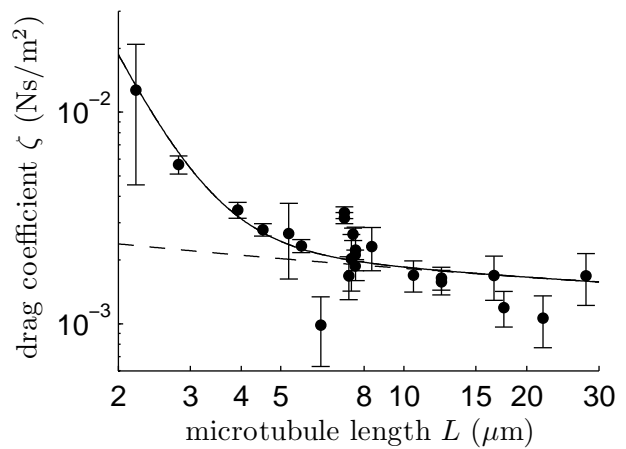}
  \caption{Drag coefficients $\zeta$ versus contour length $L$. For short
    microtubules, the experimental data lie significantly above the 
    hydrodynamic estimate $\zeta_{0}$ of Eq.~(\ref{eq:drag})
    (dashed). The solid line shows a fit of Eq.~(\ref{eq:zetafit})
    taking into account internal friction.}
  \label{fig:zeta}
\end{figure}

In the wormlike chain model, relaxation times for a filament are given by \cite{aragon}
\begin{equation}
  \label{eq:tau}
  \tau_n=\frac{\zeta L^4}{\kappa q_n^4},
\end{equation}
where $\zeta$, $L$, and $\kappa$ denote the drag coefficient per unit
length, the filament's contour length and its bending stiffness,
respectively, while $q_n$ is a mode-dependent numerical factor which
corresponds to a wave number normalized by length and is determined by the boundary
conditions. 
An estimate of $\zeta$ can be obtained from the formula for a rigid cylinder of
length $L$ and diameter $d$ held steady in a homogeneous flow in a
liquid of viscosity $\eta$ \cite{Cox}: 
\begin{equation}
  \label{eq:drag}
  \zeta_{0}=\frac{4 \pi \eta}{\ln(L/d)+2\ln(2)-1/2}.
\end{equation}
The Stokes friction coefficient for the attached bead is 
$\gamma_{\text{bead}}=6\pi\eta r \approx 1.9\times10^{-9}\,$Ns/m for $\eta \approx
10^{-3}\,$Ns/m and $r=100\,$nm being the viscosity of water and the
radius of the bead, respectively.
The drag $\zeta_{0} L$ acting on the microtubule is significantly
larger than $\gamma_ {\text{bead}}$ and only becomes
comparable for microtubules shorter than $\sim$$1\,\mu$m \footnote{Although for our shortest
  microtubule with $L=2.2\mu$m the hydrodynamic estimate
  $\zeta_0 L$ is only $\sim$3 times larger than $\gamma_{\text{bead}}$, the
  measured drag (see Fig.~\ref{fig:zeta}) is more than one order of magnitude larger than $\gamma_{\text{bead}}$.}. 
Hence, in our case, the attached bead does not need to be considered in the analysis
of the microtubule dynamics.

Due to confinement of the filament's position, the mean square
displacement of the transverse coordinate saturates
with time \cite{KroyFrey, granek}:
\begin{equation}
  \label{eq:MSD}
  \text{MSD}(t) \propto \sum _n \frac{L^3}{\kappa q_n^{4}}
  \left(1-e^{-t/\tau_n}\right) f_n(L_{\text{a}}/L) ,
\end{equation}
where $L_{\text{a}}$ refers to the contour length from the microtubule
attachment point to the bead and $f_n(L_{\text{a}}/L)$ accounts for a
modulation of the amplitude due to the position on the contour. 
In the case of one free and one clamped end, $q_n= (n-1/2)\pi$, and
the sum is taken over all modes. 
As $(q_2/q_1)^4\approx 80$, it follows that the amplitude of the first mode dominates the mean square displacement. 
Furthermore, Eq.~(\ref{eq:tau}) shows that relaxation times decrease rapidly
with mode number. 
For fluorescence microscopy studies that rely on integration times on
the order of several to tens of  milliseconds this implies that higher modes are
generally averaged over during imaging and therefore do not contribute
significantly to the measured MSD. 
Given the stiffness of microtubules, we can therefore safely assume
that our measurements essentially reflect the behavior of the first
mode. 
Contributions from the second or higher modes may occur during
measurements on long microtubules where relaxation times are
significantly longer, however these are a negligible source of error compared to the
statistical uncertainties in this length regime. 

\begin{figure}
  \centering
  \includegraphics[width=8.5cm]{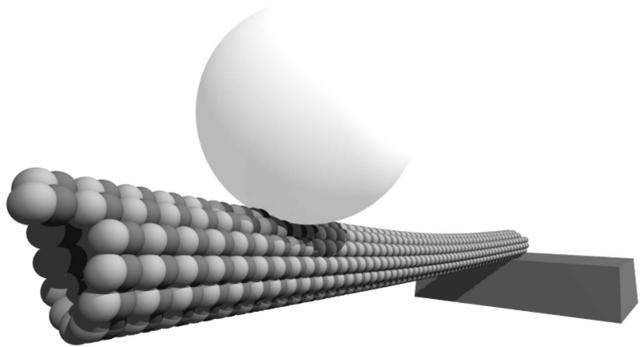}
  \caption{Schematic of the assay. One end of the microtubule is
    covalently grafted on the gold substrate. Lengths are measured
    along the contour starting at the substrate, with $L$ denoting the
    free end and $L_{\text{a}} \leq L$ marking the position of the tracer bead. 
    Different fluorescent labels for the microtubule and the bead allow for separate imaging of the two and
  prevent photobleaching. }
  \label{fig:setup}
\end{figure}

Microtubules were polymerized from unlabeled,
rhodamine-labeled and biotinilated tubulin (Cytoskeleton Inc, Denver,
CO) and stabilized in 20$\,\mu$M taxol (T1912, Sigma, St. Louis, MO)
to ensure stability of the length.
Using standard thiol chemistry (see Ref.~\cite{fra}), microtubules were
covalently grafted to a microstructured gold substrate at one end,
leaving the other end free to fluctuate in three dimensions.
A yellow-green fluorescent bead of 200$\,$nm in diameter was tightly bound to the
microtubule using specific binding between avidin and biotin. 
Fig.~\ref{fig:setup} shows a schematic of the assembly.
For 19 microtubules of lengths $L=2.2$ to $27.9\,\mu$m, $10\,000$ to $15\,000$
frames of the bead's motion were taken at frame rates of 10 to 30$\,$Hz
using integration times of 18 to 80$\,$ms.  
Two-dimensional position data were obtained from the time series 
using a custom-written Matlab (The MathWorks, Natick, MA) single
particle tracking algorithm involving Gaussian fits to the bead's image.

MSD data obtained from time averages of finite time traces are subject
to statistical errors that are nontrivial to detect and
quantify. 
In order to avoid biased results arising from correlations in the
data, error estimates on MSD data were obtained as outlined
in Ref.~\cite{flyv}. 

In addition to statistical errors, instrumentation limitations
need to be considered. 
Finite integration times average position fluctuations and thereby result in
effective low-pass filtering \cite{wong}. 
We account for the low-pass effect by extracting relaxation times from
a fit to a corrected expression for the measured MSD of first mode fluctuations
\footnote{ With only one mode being considered, the motion of a bead attached to a fluctuating microtubule can be
 modeled as that of a Brownian particle in an effective harmonic trap. 
 We hence derive a low-pass filtering correction for dynamic data building on an approach recently developed by Wong
 and Halvorsen \cite{wong} in the context of optical trapping. 
For times $t$ larger than the integration time $W$, we find 
$\text{MSD}(t) \propto  2\frac{\tau^2}{W^2}\left(e^{-W/\tau}-1+\frac{W}{\tau}
 \right)-
 \left(\frac{\sinh(\frac{W}{2\tau})}{\frac{W}{2\tau}}\right)^2e^{-t/\tau}$ .}.

\begin{figure}
  \centering
  \includegraphics{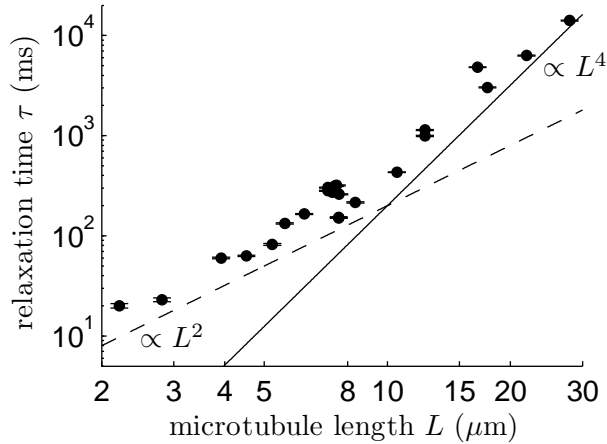}
  \caption{Relaxation times extracted from MSD. Lines of slope 4
    (solid) and 2 (dashed) are shown for comparison.}
  \label{fig:tau}
\end{figure}

In Fig.~\ref{fig:tau} the resulting relaxation time data is plotted
versus the contour length. 
For short filaments, the data approach a power law $\tau \propto L^2$
which differs considerably from the $\tau \propto L^4$ dependence
predicted by the wormlike chain model.
This observation leads to the conclusion that either the bending
stiffness $\kappa$ or the drag coefficient $\zeta$ in
Eq.~(\ref{eq:tau}) must have a hidden dependence on $L$. 

The persistence length $l_{\text{p}}=\kappa/(k_{\text{B}}T)$ can be determined from the transverse
position variance~$V$ using \cite{gholamiPRE, fra}
\begin{equation}\label{eq:V}
V=\frac{L_{\text{a}}^3}{3l_{\text{p}}} ,
\end{equation}
where a low-pass filtering corrected expression is applied \cite{wong}.
Once $l_{\text{p}}$ and hence $\kappa$ is known, the drag coefficient $\zeta$
can be computed using Eq.~(\ref{eq:tau}).

\begin{figure}
  \centering
  \includegraphics{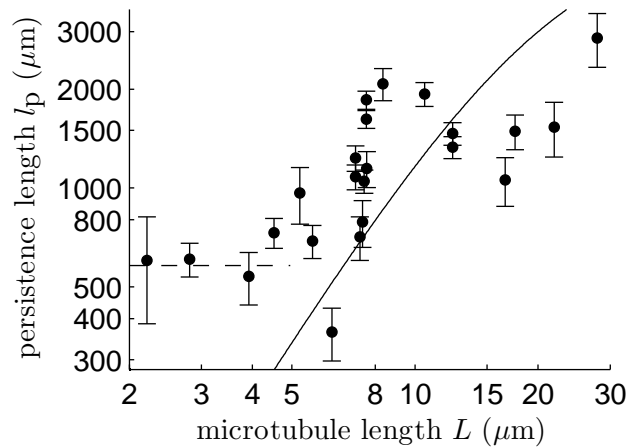}
  \caption{Persistence lengths obtained from the position
    variance. A plot of Eq.~(\ref{eq:lpfra}) using the parameters of
    $\lambda=21\,\mu$m and $l_{\text{p}}^{\infty}=6300\,\mu$m obtained by
    Ref.\cite{fra} is shown for comparison (solid line). For short
    microtubules, the data deviate from the Timoshenko theory and level to
  a plateau (dashed).}
  \label{fig:lp}
\end{figure}

The plots of $l_{\text{p}}$ and $\zeta$ versus $L$ in Figs.~\ref{fig:lp}
and~\ref{fig:zeta} immediately reveal that both quantities exhibit unexpected
behavior. 
The persistence length shows a significant increase with overall
filament length, similar to the length-dependence obtained in
Ref.~\cite{fra} where, based on the Timoshenko beam formalism
\cite{timo} that describes macroscopic elastic beams with shear contributions, the expression 
\begin{equation}
   \label{eq:lpfra}
   l_{\text{p}}= \frac{l_{\text{p}}^{\infty}}{1+\left(\frac{\lambda}{L}\right)^2}
\end{equation}
was derived. The parameter $\lambda$ accomodates for decreases in microtubule stiffness due
to shear contributions from inter-protofilament sliding.
In our case however, a fit of Eq.~(\ref{eq:lpfra}) cannot provide much
information as our data range does not include the saturation at $l_{\text{p}}^{\infty}$
predicted for very long microtubules. 
While, for microtubules longer than $\sim$$5\,\mu$m, qualitative
agreement of the data with the quadratically increasing regime of
Eq.~(\ref{eq:lpfra}) is evident, shorter
microtubules seem to deviate from this behavior and level to a
plateau. 
Averaging the values for the three shortest microtubules yields a
value of $580\pm100\,\mu$m.
Interestingly, the existence of such a plateau value is predicted by
the more general theory of wormlike bundles \cite{clausPRL, bathe} that takes
into account the microscopic filament architecture.
The Timoshenko theory derives an effective bending rigidity from
stretch and shear terms, which result in fully coupled bending for
long microtubules and an intermediate, shear-dominated length regime
where rigidity scales with $L^2$. 
In addition to these terms, the constituent filaments in a wormlike bundle have a bending
stiffness that dominates the rigidity of the decoupled bundle in the short
length regime.

At the same time, the drag coefficients shown in Fig.~\ref{fig:zeta}
significantly deviate from the hydrodynamic estimate of
Eq.~(\ref{eq:drag}) for short microtubules. 
The sharp increase in this length regime is consistent with the presence of internal friction as introduced
by Poirier and Marko \cite{Poirier}, who derived an additional friction
term of
\begin{equation}\label{eq:intf}
\zeta_{\text{int}}=\varepsilon \left(\frac{q_n}{L}\right)^4 
\end{equation}
to accomodate for dissipation during conformational changes or due to
liquid flowing through narrow pores in a biofilament. 
In the case of microtubules, both contributions seem reasonable due to
the complex composite architecture and the large fluid cylinder inside
the filament.
The parameter $\varepsilon$ is proportional to an effective viscosity
summarizing the friction losses inside the filament. 
A fit of the expression 
\begin{equation}\label{eq:zetafit}
\zeta= \zeta_{0} + \zeta_{\text{int}}=\frac{4 \pi \eta}{\ln(L/d)+2\ln(2)-1/2}+\varepsilon \left(\frac{q_1}{L}\right)^4
\end{equation}
with $q_1=\pi/2$ yields a coefficient $\varepsilon=4.3\pm0.5
\times 10^{-26}$~Nsm$^2$ which is more than one order of magnitude
smaller than the numbers of 1.6 $\times 10^{-24}$~Nsm$^2$ and $6.9
\times 10^{-25}$~Nsm$^2$ obtained by Brangwynne et al.~\cite{bend} and Janson
and Dogterom \cite{Janson}, respectively.
This discrepancy suggests that high mode deformations of long
microtubules may be governed by different dissipative mechanisms than first mode
deformations of short microtubules.

In this letter, we presented a first analysis of relaxation times and
drag coefficients for microtubules in the
length range most relevant to essential cellular functions such as
cell division.
Our results show that the dynamic behavior for this short length
regime significantly deviates from predictions of the wormlike chain
model. 
The dynamics seem to be modulated by a length-dependent persistence
length as well as additional friction contributions that may result from internal dissipation.
While microtubules longer than $\sim$$5\,\mu$m display a persistence length behavior
consistent with the Timoshenko model, shorter microtubules show a
uniform persistence length of $580\pm100\,\mu$m which is consistent
with decoupled bending of constituent protofilaments as predicted by
the theory of wormlike bundles \cite{clausPRL, bathe}.
These findings not only emphasize the importance of molecular architecture to
be taken into account for models of semiflexible polymer mechanics,
but also suggest a wealth of possibilities for natural modulation of
biopolymer properties in cells.

\begin{acknowledgments}
The authors would like to thank Claus Heussinger for helpful
discussions. 
K.M.T. and F.P. gratefully acknowledge support by the German
National Academic Foundation and the Landesstiftung Baden-W\"urttemberg (Forschungsprogramm
Optische Technologien), respectively.
E.F. acknowledges support from the DFG through SFB 486, and from the
German Excellence Initiative via the NIM program.
This research was supported by NSF Grant No. CMMI-0728166.
\end{acknowledgments}

\bibliography{ref_prl}

\end{document}